\documentclass[aps,pre,showpacs,twocolumn,superscriptaddress,letterpaper]{revtex4}
\usepackage{mathrsfs}
\usepackage{amsmath}
\usepackage{bm}
\usepackage{color}
\usepackage{graphicx}
\usepackage{hyperref}
\hypersetup{pdfpagemode=UseNone,pdfstartview=FitH,pdfstartpage=1}

\begin{document}


\title{Comparisons and Applications of Four Independent Numerical Approaches for Linear Gyrokinetic Drift Modes
 }
\author{H. S. Xie}
\email[]{Email: huashengxie@gmail.com} \affiliation{Fusion
Simulation Center, State Key Laboratory of Nuclear Physics and
Technology, School of Physics, Peking University, Beijing 100871,
China}
\author{Y. Y. Li}
\affiliation{Institute for Fusion Theory and Simulation, Department
of Physics, Zhejiang University, Hangzhou 310027, China}
\author{Z. X. Lu\footnote{Current address: Max-Planck-Institut f\"ur Plasmaphysik, Boltzmannstr. 2,
85748 Garching, Germany.}} \affiliation{Fusion Simulation Center,
State Key Laboratory of Nuclear Physics and Technology, School of
Physics, Peking University, Beijing 100871, China}
\author{W. K. Ou}
\affiliation{Fusion Simulation Center, State Key Laboratory of
Nuclear Physics and Technology, School of Physics, Peking
University, Beijing 100871, China}
\author{B. Li}
\email[]{Email: bli@pku.edu.cn} \affiliation{Fusion Simulation
Center, State Key Laboratory of Nuclear Physics and Technology,
School of Physics, Peking University, Beijing 100871, China}

\date{\today}

\begin{abstract}
To help reveal the complete picture of linear kinetic drift modes,
four independent numerical approaches, based on integral equation,
Euler initial value simulation, Euler matrix eigenvalue solution and
Lagrangian particle simulation, respectively, are used to solve the
linear gyrokinetic electrostatic drift modes equation in Z-pinch
with slab simplification and in tokamak with ballooning space
coordinate. We identify that these approaches can yield the same
solution with the difference smaller than 1\%, and the discrepancies
mainly come from the numerical convergence, which is the first
detailed benchmark of four independent numerical approaches for
gyrokinetic linear drift modes. Using these approaches, we find that
the entropy mode and interchange mode are on the same branch in
Z-pinch, and the entropy mode can have both electron and ion
branches. And, at strong gradient, more than one eigenstate of the
ion temperature gradient mode (ITG) can be unstable and the most
unstable one can be on non-ground eigenstates. The propagation of
ITGs from ion to electron diamagnetic direction at strong gradient
is also observed, which implies that the propagation direction is
not a decisive criterion for the experimental diagnosis of turbulent
mode at the edge plasmas.
\end{abstract}

\pacs{52.35.Py, 52.30.Gz, 52.35.Kt}

\maketitle

\section{Introduction and motivation}\label{sec:intro}
Drift modes or instabilities are micro-instabilities, driven by
equilibrium non-uniformity, such as temperature and density
gradients, and magnetic field non-uniformity. Those non-uniformities
can lead to various particle drifts (e.g., gradient drift, curvature
drift), and these particle drifts may cause particular eigen
oscillation or instability, and thus these corresponding modes are
called drift modes. Drift wave turbulence is considered to be the
major cause for anomalous transport \cite{Horton1999}, which is an
active research field in plasma physics. Thus, it is important to
understand drift modes.

Gyrokinetic theory \cite{Brizard2007}, more accurate than fluid
theory, is a major tool to study the low frequency
($\omega\ll\Omega_{ci}$, where $\Omega_{ci}$ is ion cyclotron
frequency) drift modes. In principle, different approaches to solve
the same model should give the same solution. We noticed that the
discrepancy between different numerical solvers can be larger than
$10\%$  (cf. Ref.\cite{Kotschenreuther1995,Rewoldt2007}) for weak
gradient core plasma and even larger than $50\%$ for strong gradient
edge plasma (cf. Ref.\cite{Wang2012}) in literature. We have also
noticed a recent carefully verification of global gyrokinetic
tokamak Euler and particle codes in Ref.\cite{Tronko2017}, where two
codes GENE and ORB5 yield similar gyrokinetic models but still have
visible (around 5\%) deviations in real frequency and growth rate.
To reveal the reasons (e.g., model difference, boundary condition
difference, grids convergence, limitation of numerical approach) for
this discrepancy is crucial to quantitatively study the drift modes.
It also seems that no detailed benchmarks of four different
approaches (based on integral equation, Euler initial value
simulation, Euler eigenvalue solution and Lagrangian particle
simulation, respectively) for the same gyrokinetic model are found
in the literature.  In this work, we use exactly the same model
equation for all four approaches, in contrast to many inter-code
benchmarks where both the numerics and the model equations differ to
some degree. Thus, we hope this work can also be useful as a
reference base, and help other researchers to choose which approach
to use for their own purposes in consideration of the balance of
computational time and accuracy. We have implemented all the four
approaches to the code series MGK\footnote{Matlab version codes can
be found at: \url{http://hsxie.me/codes/mgk}.} (Multi-approach
GyroKinetic code).

We also find that it is very useful to understand the distribution
of the linear solutions by cross checking the solutions of each
approach. For example, using these four independent approaches, we
apply them to reveal the relation between the interchange mode and
entropy mode in Z-pinch\cite{Ricci2006} and find that they are on
the same branch and only one unstable mode exists in the system for
small $\eta$, with $\eta=L_n/L_T$ the ratio of temperature gradient
to density gradient. These approaches are also used to study the
multi-eigenstates of ion temperature gradient modes coexisting in
strong gradient at tokamak edge.

In the following sections, Sec.\ref{sec:model} gives the model
linearized gyrokinetic equation. Sec.\ref{sec:0d} gives the details
of four approaches to solve the kinetic ion and electron
zero-dimensional (0D) model in Z-pinch. Sec.\ref{sec:1d} extends the
0D approaches to one-dimensional (1D) model to study the ion
temperature gradient mode in tokamak using kinetic ion but adiabatic
electron with ballooning coordinate. Sec.\ref{sec:gradient} studies
the drift modes under strong gradient. Sec.\ref{sec:summ} summarizes
the present study.

\section{Gyrokinetic Linear Model}\label{sec:model}
Gyrokinetic model\cite{Antonsen1980,Chen1991,Brizard2007} can
describe the low frequency physics accurately under the assumptions:
$\omega/\Omega_{ci}\sim\rho_i/L\sim
k_\parallel/k_\perp\sim\delta\ll1$. Assuming Maxwellian equilibrium
distribution function $F_{0}=n_0F_M$, $F_M=(\frac{m}{2\pi
T})^{3/2}e^{-m\epsilon/T}$, where $\epsilon=v^2/2$,
$v^2=v_\parallel^2+v_\perp^2$, $\mu=v_\perp^2/2B$, the perturbed
distribution function after gyrophase average is
\begin{equation}\label{eq:gk_f0}
  f_{\alpha} = \frac{q_\alpha}{m_\alpha} \phi \frac{\partial F^{\alpha}_{0}}{\partial \epsilon} + J_{0}(k_\perp\rho_{\alpha})
  h_{\alpha},
\end{equation}
where the Bessel function of the first kind $J_0$ comes from
gyrophase average, the non-adiabatic linearized gyrokinetic response
$h_\alpha$ satisfies
\begin{equation}\label{eq:gk_dh0}
(\omega - \omega_{D\alpha} + i v_{\parallel} {\bf b} \cdot {\bf
\nabla}) h_{\alpha}
  = - (\omega - \omega_{*\alpha}^T)
  \frac{\partial F^{\alpha}_{0}}{\partial \epsilon} \frac{q_\alpha}{m_\alpha}J_{0} \phi.
\end{equation}
Here the collision operator is neglected, and the parameters are
\begin{eqnarray*}
  \rho_{\alpha} &=& \frac{v_{\perp}}{\Omega_{\alpha}}, ~~~~\Omega_\alpha=\frac{q_\alpha B}{m_\alpha c}, \\
  \frac{\partial F^{\alpha}_{0}}{\partial \epsilon}&=&-\frac{m_\alpha F_0^\alpha}{T_\alpha}, \\
  \omega_{*\alpha}^T &=& \frac{{\bf k}_{\perp} \times {\bf b}
\cdot {\bf \nabla} F^{\alpha}_{0} }{ -\Omega_{\alpha}F^{\alpha}_{0\epsilon}}, \\
  \omega_{D\alpha} &=& {\bf k}_{\perp} \cdot {\bf v}_d={\bf k}_{\perp} \cdot {\bf b} \times \frac{\mu \nabla B
+ v_{\parallel}^{2} {\bf b} \cdot \nabla {\bf b}}{\Omega_{\alpha}}, \\
  {\bf b} &=& {\bf B} / B,
\end{eqnarray*}
$\alpha=i,e$ represents particle species. Here, ${\bf B}$ is the
magnetic field, and $q_\alpha$, $m_\alpha$, $T_\alpha$,
$\Omega_{\alpha}$, $\rho_{\alpha}$, $\omega_{*\alpha}^T$ and
$\omega_{D\alpha}$ are the charge, mass, temperature, cyclotron
frequency, gyroradius, diamagnetic drift frequency and magnetic
(gradient and curvature) drift frequency for the species $\alpha$,
respectively. In electrostatic case, the quasi-neutrality condition
(Poisson equation)
\begin{equation}\label{eq:gk_n}
  \sum_\alpha q_\alpha\int f_{\alpha} d^{3} v =0.
\end{equation}
is used for field equation, where the notation for velocity integral
$\int d^3 v\equiv
2\pi\int_{-\infty}^{\infty}dv_\parallel\int_{0}^{\infty}v_\perp
dv_\perp$, and the gyrophase average yielded $J_0$ has been
contained in Eq.(\ref{eq:gk_f0}). In this work, we will treat both
zero-dimensional (0D) model for Z-pinch and one-dimensional (1D,
along field line) model for tokamak. For 1D, ${\bf b} \cdot {\bf
\nabla} =
\partial_{l}$; for 0D, ${\bf b} \cdot {\bf \nabla} = ik_\parallel$.

\section{Outline of Four approaches}\label{sec:0d}

In Z-pinch, $\omega_{*\alpha}=-\frac{cT_{0\alpha}}{q_\alpha B
n_0}\nabla n_0\cdot{\bm b}\times{\bm
k}=k\rho_\alpha\frac{v_{t\alpha}}{L_n}$,
$\omega_{d\alpha}=\frac{v_{t\alpha}^2}{\Omega_{c\alpha}B^2}{\bm
B}\times\nabla B\cdot{\bm
k}=k\rho_\alpha\frac{v_{t\alpha}}{R}=\omega_{*\alpha}L_n/R$, where
$L_n^{-1}=-d\ln n_0/dr$, ${\bm b}$ being the unit vector of the
magnetic field $\bm B$, and $q_{\alpha}$,
$v_{t\alpha}=\sqrt{T_{0\alpha}/m_\alpha}$,
$\Omega_{c\alpha}=q_\alpha B_\theta/(m_\alpha c)$, $m_\alpha$, and
$\rho_{c\alpha}=v_{t\alpha}/\Omega_{c\alpha}$ being the charge,
thermal velocity, cyclotron frequency, mass and gyroradius for the
species $\alpha$, respectively (note $\rho_e<0$),
$\tau=\tau_e=T_{0e}/T_{0i}$. For $b\ll1$, the Bessel function
$J_0^2(b)=1-b^2/2$. We will treat $q_i=-q_e=e$ by default. For
$\tau=1$, $\omega_*=\omega_{*i}=-\omega_{*e}$ and
$\omega_d=\omega_{di}=-\omega_{de}$. We also define $\kappa_n\equiv
R/L_n$, $\kappa_T\equiv R/L_T$, $\epsilon_n\equiv L_n/R$ and
$\eta=\kappa_T/\kappa_n$, with $L_T=(-d\ln T/dr)^{-1}$. One should
note that: In this section, the ion diamagnetic drift frequency
$\omega_{di}>0$, to be consistent with Ref.\cite{Ricci2006};
Whereas, in the next section to study tokamak ITG mode,
$\omega_{di}<0$, to be consistent with standard tokamak community
notation, e.g., Ref.\cite{Dong1992}.

\begin{figure}
 \centering
  \includegraphics[width=8.5cm]{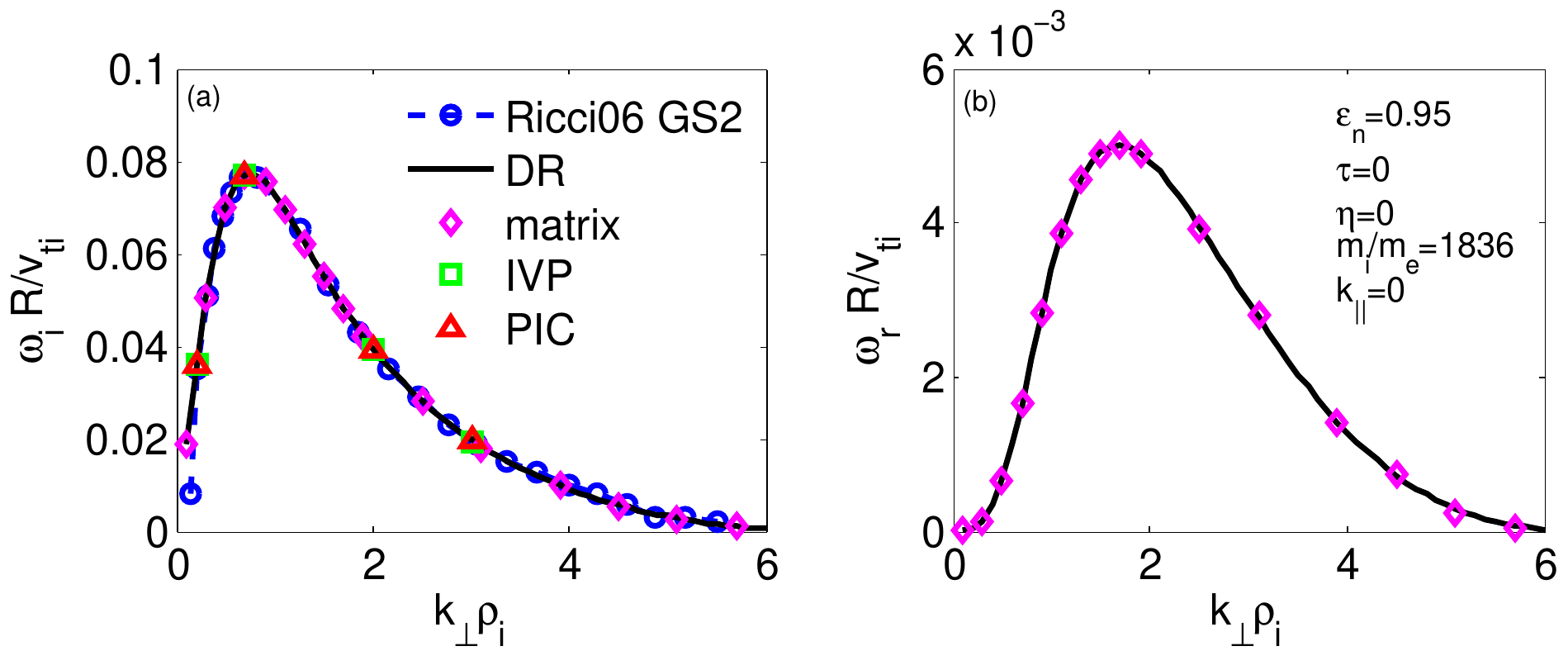}\\
  \caption{Benchmark and comparison of four approaches for entropy mode.}\label{fig:cmp_ricci06_fig4}
\end{figure}

\subsection{Integral Dispersion relation}
In the 0D slab limit, the gyrokinetic model Eqs.(\ref{eq:gk_dh0})
and (\ref{eq:gk_n}) can readily be solved and yield the following
integral dispersion relation
\begin{equation}\label{eq:dr_0d}
  \sum_\alpha \frac{q_\alpha^2}{T_\alpha}\Big\{1-\int \frac{(\omega -
  \omega_{*\alpha}^T)}{(\omega - \omega_{D\alpha} -
  k_\parallel v_{\parallel})}J_0^2F_{M\alpha} d^{3} v \Big\}=0,
\end{equation}
where all particle species (ion and electron) are treated by
gyrokinetic model. Eq.(\ref{eq:dr_0d}) is an integral equation.
Without analytical continuation to $Im(\omega)\leq0$, the above
equation can only represent the $Im(\omega)>0$ growing modes
correctly. The double integration $dv_\parallel dv_\perp$ can be
evaluated by adaptive Simpson method. Using spectral method for
$dv_\parallel$ \cite{Weideman1994,Xie2013} and Gauss-Kronrod method
for $dv_\perp$ \cite{Gurcan2014}, the above double integration can
be evaluated $\sim200$ times faster. In practical test, we find that
the Gauss method is difficult to obtain high accuracy for
$\omega\to0$ and thus is not accurate to study some parameters
(especially for $k_\perp\rho_i\ll1$ and $\gg1$) for the entropy mode
in subsection \ref{sec:bench_0d}. Hence, we mainly use adaptive
Simpson method in this work, which is easily to control the
computation precision.

Define $y=v/v_{t\alpha}$, the integral Eq.(\ref{eq:dr_0d}) is
normalized to{\scriptsize
\begin{equation}\label{eq:dr_0d_1}
  \sum_{\alpha=e,i} \frac{q_\alpha^2}{T_\alpha}\Big\{1-\frac{1}{\sqrt{2\pi}}\int \frac{(\omega -
  \omega_{*\alpha}^T)J_0^2(k_\alpha y_\perp)}{(\omega - \omega_{D\alpha} -
  k_{z\alpha} v_{\parallel})}e^{-\frac{y^2}{2}} y_\perp dy_\perp dy_\parallel \Big\}=0,
\end{equation}}
with
$\omega_{*\alpha}^T=\omega_{d\alpha}[\kappa_n+\kappa_T(y^2/2-3/2)]$,
$\omega_{D\alpha}=\omega_{d\alpha}(y_\parallel^2+y_\perp^2/2)$,
$v_{te}=v_{ti}\sqrt{\tau m_i/m_e}$,
$\Omega_e=\frac{q_em_i}{q_im_e}\Omega_i$,
$\rho_e=\rho_i\frac{q_i}{q_e}\sqrt{\tau\frac{m_e}{m_i}}$,
$k_\alpha=k_\perp\rho_\alpha$, $k_{z\alpha}=k_\parallel
v_{t\alpha}$. Normalized by $\omega_0=v_0/R_0$, with $v_0=v_{ti}$
and $R_0=R$. Thus, $\omega_{di}=k_\perp$, $\omega_{de}=k_\perp
\tau\frac{q_i}{q_e}$, $k_i=k_\perp$,
$k_e=k_\perp\frac{q_i}{q_e}\sqrt{\tau\frac{m_e}{m_i}}$,
$k_{zi}=k_\parallel R$ and $k_{ze}=k_\parallel R\sqrt{\tau
m_i/m_e}$. The dispersion relation can be solved by standard root
finding approach.

\subsection{Euler initial value problem}
Define
$g_\alpha(v_\parallel,v_\perp)=h_\alpha-\frac{q_\alpha}{T_\alpha}F_0^\alpha
J_0\phi$, and using $\omega=i\partial_t$, we can transform the
original equation to
\begin{eqnarray}\label{eq:ivp_0d}
  \partial_t g_\alpha &=& -i(\omega_{D\alpha}+k_{z\alpha}y_\parallel)g_\alpha-
  i[\omega_{D\alpha}-\omega_{T\alpha}]\phi J_{0\alpha}e^{-\frac{y^2}{2}}, \\\label{eq:ivp_0d_phi}
  \phi &=&
  \frac{c}{\sqrt{2\pi}}\int\Big(g_iJ_{0i}+\frac{g_eJ_{0e}}{\tau}\Big)y_\perp
  dy_\perp dy_\parallel,
\end{eqnarray}
where $c=\frac{1}{[1-\Gamma_{0i}+(1-\Gamma_{0e})/\tau]}$,
$\omega_{T\alpha}=\omega^T_{*\alpha}$, $J_{0\alpha}=J_0(k_\alpha
y_\perp)$, $\Gamma_{0\alpha}=I_0(b_\alpha)e^{-b_\alpha}$,
$b_\alpha=k_\perp^2\rho_\alpha^2=k_\alpha^2$.

The above equation can be solved as initial value problem via Euler
discretization of the velocity coordinate $(v_\parallel,v_\perp)$.
We also notice that the electron and ion can be treated in the same
time scale in the above equations, since their velocities can be
normalized by their own thermal velocities, i.e., $y=v/v_{t\alpha}$.
This only happens when the parallel and perpendicular dynamics can
be treated via the parameters $k_\parallel$ and $k_\perp$.

\subsection{Euler eigenvalue problem}
Eqs.(\ref{eq:ivp_0d}) and (\ref{eq:ivp_0d_phi}) can also be solved
as eigenvalue problem using the same Euler discretization as an
initial value problem, via $\partial_t=-i\omega$, $\lambda X=AX$,
$X=[g_i^{n_{y_\parallel}\times
n_{y_\perp}},g_e^{n_{y_\parallel}\times n_{y_\perp}}, \phi]^T$. To
keep the matrix sparse, we have done a further time derivative
$\partial_t$ on the field equation (\ref{eq:ivp_0d_phi}). This is
similar to the latter 1D case Eq.(\ref{eq:itg1d3_2}).

We notice that this matrix eigenvalue approach may be the best
approach for the present model, considering that it can give all the
solutions in the system at the same time and thus will not miss
solutions. This approach can be both accurate and fast for matrix
dimension $N=2\times n_{y_\parallel}\times n_{y_\perp}+1\leq5000$.
Numerical results will be shown later. It should be noticed that
only direct matrix solvers can give all the eigenmodes of the
system. However, their execution time and memory scale with
$O(N^p)$, where usually $2.0<p\sim2.6<3.0$. For more realistic
systems (e.g. edge tokamak cases with large resolution
requirements), such solvers therefore quickly become too expensive
to use, and have to be replaced by iterative solvers, which will
return a limited number of solutions according to a certain
selection criterion. Thus, in more realistic systems, we can use low
resolution to obtain all solutions and which can show the
distributions of the solutions in the $\omega_i$ vs. $\omega_r$
complex plane. Then, we use the rough solution (e.g., we may be only
interested in the most unstable solutions) as initial guess to
obtain high resolution solution(s) via sparse matrix iterative
approach, e.g., the eigs() function in MATLAB.

\subsection{Particle simulation}
The above model can also be solved using $\delta f$ approach
\cite{Parker1993}: defining particle weight $w_\alpha=g_\alpha/F_0$,
$F_0=e^{-y^2/2}$, initial loading $g_\alpha(y_\parallel,y_\perp,w)$
with Gaussian random number $y_{\parallel j}=randn(n_p,1)$,
$y_{\perp j}=\sqrt{randn(n_p,1)^2+randn(n_p,1)^2}$, and with weight
being a small value, e.g., $w_j=10^{-6}$. Here, $n_p$ is particle
number, $j=1,2,\cdots,n_p$ is particle index, and $randn()$
generates normal distribution $F_0=\frac{1}{\sqrt{2\pi}}e^{-x^2/2}$.
The particle simulation evolution equations are
\begin{eqnarray}
  \dot{w}_j &=& -i(\omega_{D\alpha}+k_{z\alpha}y_{\parallel j})w_j-i[\omega_{D\alpha}-\omega_{T\alpha}]\phi J_0(k_\alpha y_\perp), \\
  \phi &=& c\sum_j\Big(w_{ij}J_{0i}+w_{ej}J_{0e}/\tau\Big),
\end{eqnarray}
$\dot{y}_{\parallel j}=0$ and $\dot{y}_{\perp j}=0$.

This particle simulation approach can be seen as Monte-Carlo method,
and one can refer to Ref.\cite{Aydemir1994} for the general theory
and validity of it. Since this approach is based on Monte-Carlo
sample, the noise level is $\sim1/\sqrt{n_p}$. Thus, if we want the
relative error decreasing from $10\%$ to $1\%$, we need
$n'_p=100n_p$. In this viewpoint, particle approach is not a good
choice to obtain high accuracy. The benefit is that this approach is
straightforward and easily coding and parallelizing.

\begin{widetext}

\begin{table}[!h]
\tabcolsep 5mm \caption{To converge to $\omega_i=0.0770$
($\omega_r=0.0017$) for $k_\perp\rho_i=0.7$ with error less than
$1\%$, the typical grids and computation times
required.}\label{tab:mgk0d_1}
\begin{center}
\begin{tabular}{c|cc}
\hline\hline Approach & Grids & Typical cputime\\\hline
DR & accurate to $10^{-6}$ & $<$1s \\
matrix & $n_{v_\parallel}=64$, $n_{v_\perp}=64$ & $\sim$1s \\
ivp & $n_{v_\parallel}=64$, $n_{v_\perp}=64$, $\Delta t=0.01$, $n_t=10^4$ & $\sim$5min\\
pic & $n_p=4\times10^5$, $\Delta t=0.01$, $n_t=10^4$ & $\sim$20min
\\\hline\hline
\end{tabular}
\end{center}
\end{table}

\begin{figure}
 \centering
  \includegraphics[width=14.5cm]{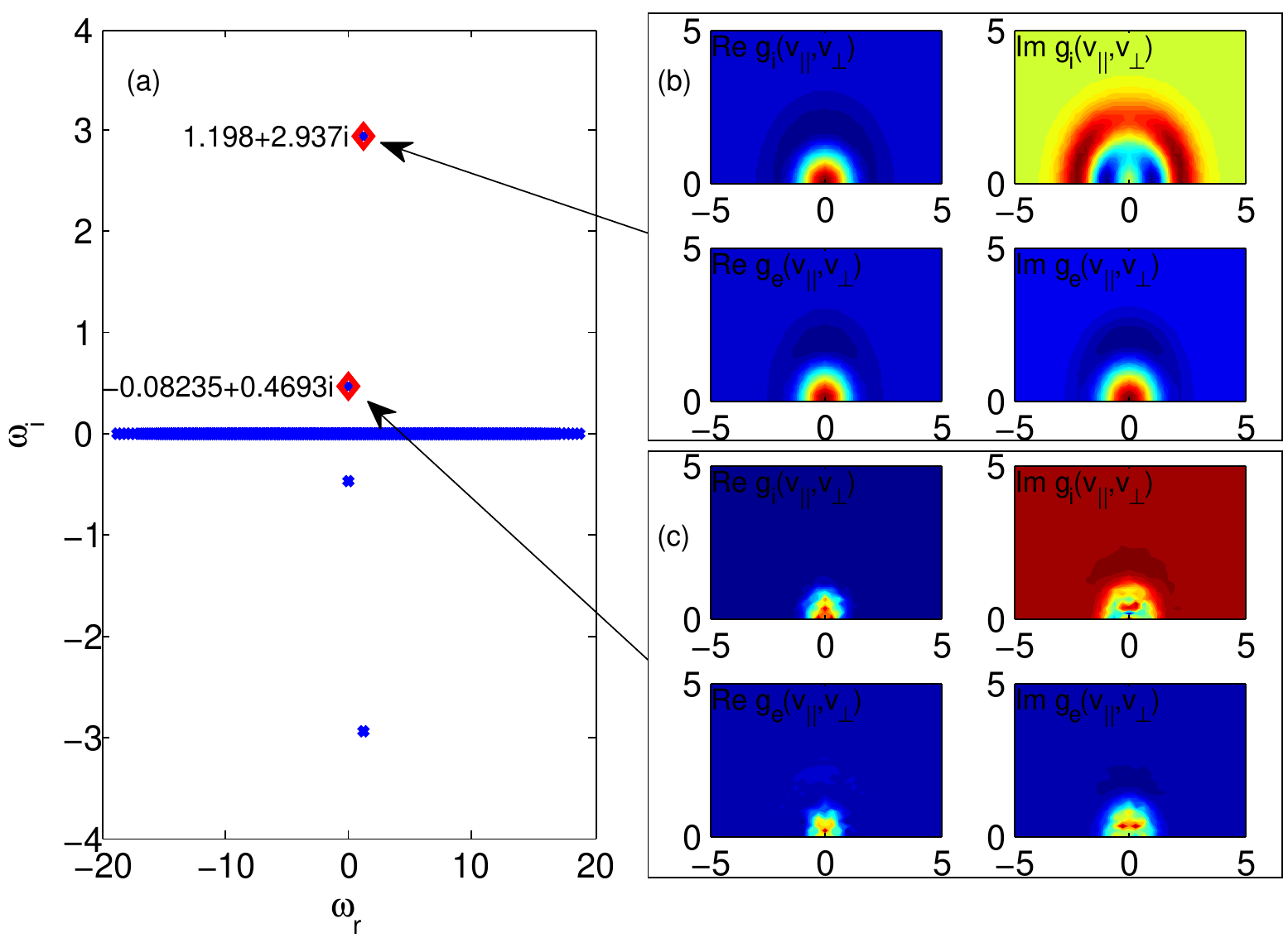}\\
  \caption{ (a) The matrix method can give all solutions in the system.
  Two unstable modes exist under parameters
  $\eta=1.5$, $k_\perp\rho_i=0.5$, $\epsilon_n=0.3$, and the corresponding mode
  structures in velocity space are shown in (b) and (c), with
  $n_{v_\parallel}=n_{v_\perp}=32$. By refining the grids,
  the accurate values of these two solutions are $\omega=1.199+2.936i$
  and $-0.019+0.471i$, which are also verified by DR method.}\label{fig:mgk0d_matrix_eta2}
\end{figure}

\begin{figure}
 \centering
  \includegraphics[width=13.5cm]{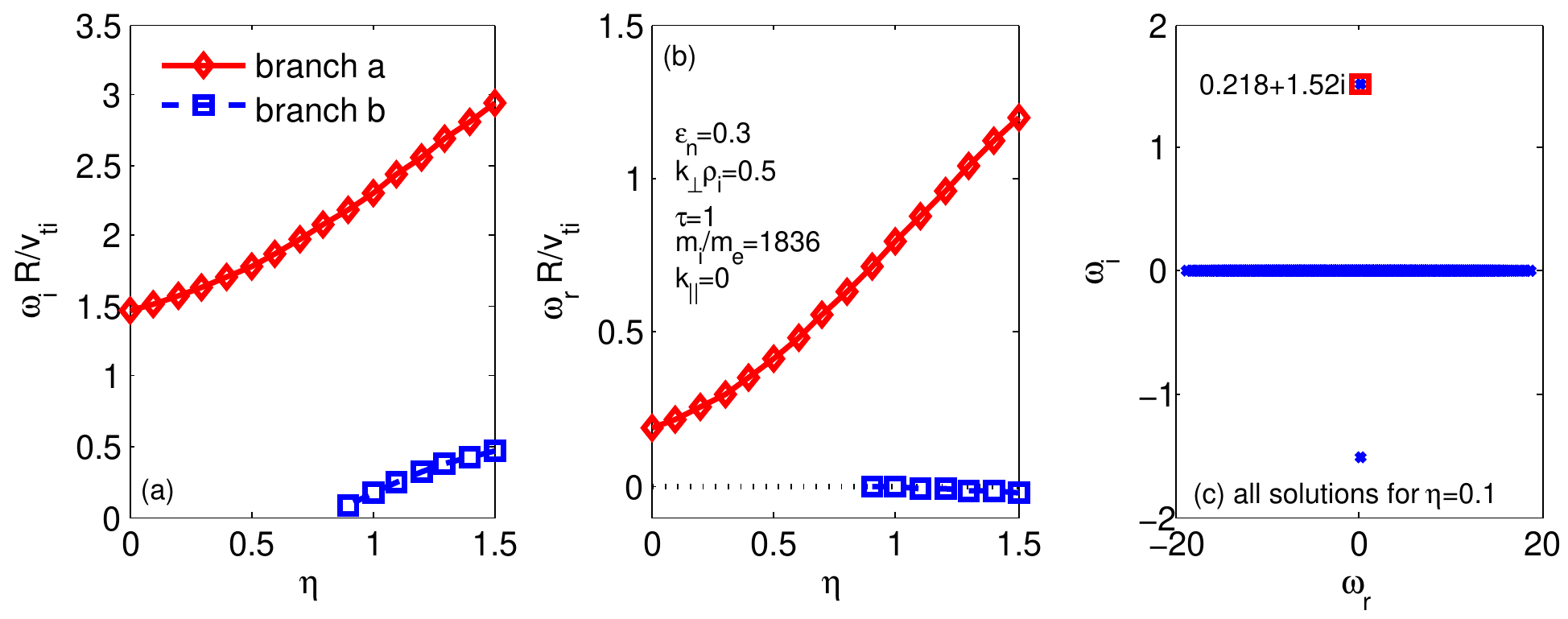}\\
  \caption{Scan $\eta$ for entropy mode under parameters $k_\perp\rho_i=0.5$, $\epsilon_n=0.3$,
  which shows that at $\eta>0.9$, two modes are unstable;
  whereas only one unstable mode for small $\eta$. (a) and (b) are obtained from DR method.
  (c) is from matrix method, which verifies that for $\eta=0.1$, only one unstable mode exists.}\label{fig:mgk0d_dr_scan_eta}
\end{figure}

\begin{figure}
 \centering
  \includegraphics[width=13.5cm]{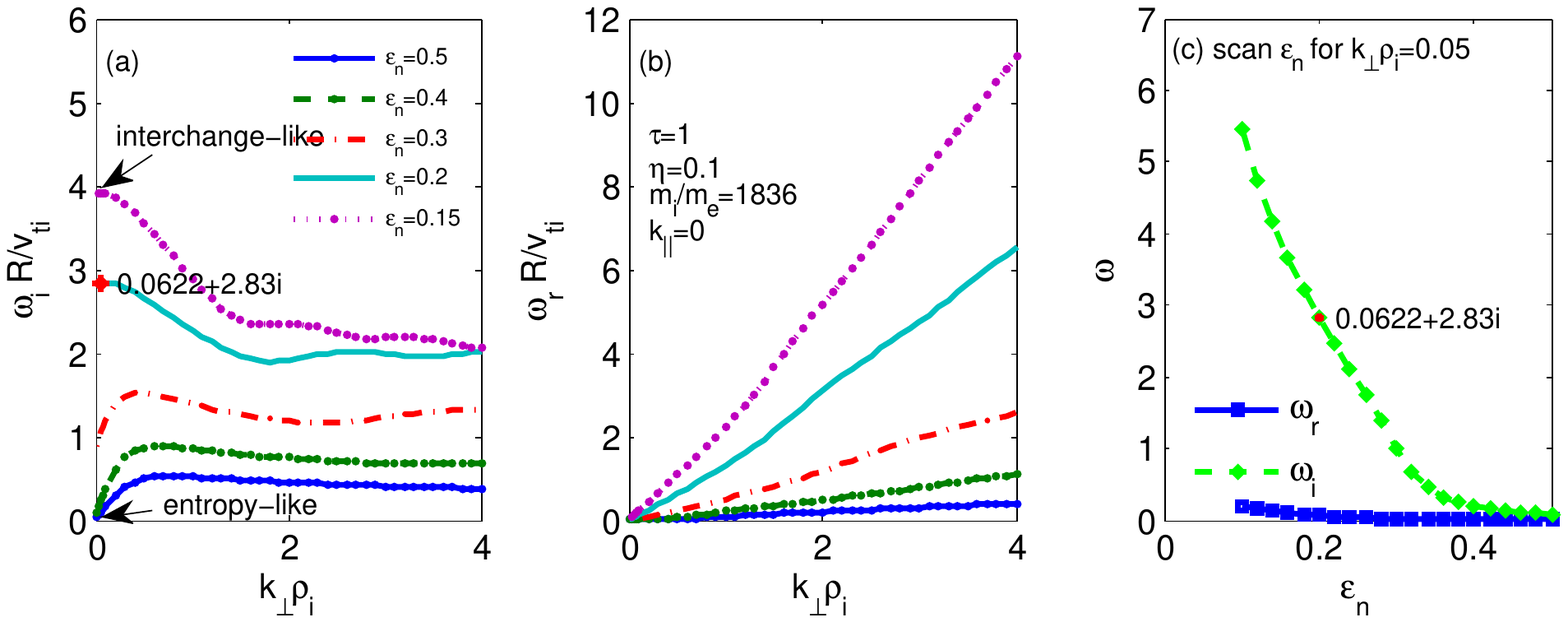}\\
  \caption{The entropy-like mode and interchange-like mode is on the same branch via scanning $\epsilon_n$.}\label{fig:mgk0d_scan_epsn}
\end{figure}

\end{widetext}

\subsection{Benchmark, comparisons of four approaches and
applications}\label{sec:bench_0d}

In this section, we use the mgk0d version. In the present study, all
four approaches are implemented using MATLAB, and run with single
core. We firstly benchmark our four approaches with the GS2 results
of Fig. 4 of Ref.\cite{Ricci2006} for the entropy mode, with
parameters $\eta=0$, $\tau=1$, $k_\parallel=0$, $\epsilon_n=0.95$
and $m_i/m_e=1836$. Here, entropy mode is one type of drift mode
which can be unstable when macro-scale interchange mode is stable
\cite{Ricci2006}. Fig.\ref{fig:cmp_ricci06_fig4} shows that our four
approaches can be almost identical to each other, with error less
than $1\%$. If we refine the grids, they can be even more close. The
result is also close to Ref.\cite{Ricci2006}. Hence, we can conclude
that all these four approaches can correctly study the linear
physics of the gyrokinetic drift modes.

For the $k_\perp\rho_i=0.2$ case, the typical grid numbers and
computation time required for each approach are listed at Table
\ref{tab:mgk0d_1}, where we find the most accurate and fast approach
is the integral dispersion relation (DR) method, and the matrix
eigenvalue approach (matrix) is also very fast and accurate. The
Euler initial value (ivp) and particle (PIC) method are slow for
this case, because large number of time steps $n_t$ is required to
diagnose the complex frequency accurately. For $k_\perp\rho_i>1.5$,
the grids need at least double to make sure the error less than
$1\%$, and thus the computation times for both approaches are
usually larger than 1hr.

Dispersion relation can only give the complex frequency $\omega$,
whereas the other three approaches can also give the structure of
$g(v_\parallel,v_\perp)$ directly, which can help to show the
resonance structure in the phase space. To calculate the mode
structure of $g(v_\parallel,v_\perp)$ in DR method, one needs to
plug the solution back into the equation for the distribution
function. Particle simulation and initial value simulation can also
easily study the most weakly damped mode (see next section for 1D
case). The matrix method can not represent the damping mode
correctly. We will show these features step by step. Since DR and
matrix methods are both fast and accurate, we will mainly use these
two approaches for the following studies.

The purpose of the present work is not merely to show that four
approaches can give the same solution, but to help to reveal the
unclear physics when single approach is not sufficient to draw a
conclusion. Now, we begin to show a more complete picture of entropy
mode, i.e., how many unstable solutions exist, ion mode or electron
mode, what is the relationship with interchange mode.

Firstly, we show all solutions in the system via matrix eigenvalue
method. For $\eta=1.5$, $\tau=1$, $k_\parallel=0$, $\epsilon_n=0.3$
$k_\perp\rho_i=0.5$ and $m_i/m_e=1836$, all solution in the system
is shown in Fig.\ref{fig:mgk0d_matrix_eta2}a. We find that only two
unstable modes exist and one is in ion diamagnetic direction and one
is in electron diamagnetic direction, i.e., there exist two types of
entropy mode. And we call them ion mode and electron mode
respectively. In Fig.\ref{fig:mgk0d_matrix_eta2}b and c, we can find
that the phase space structure is wider for ion mode and narrow for
electron mode by comparison with their thermal velocities.
Fig.\ref{fig:mgk0d_dr_scan_eta}a and b shows more clearly of the two
branches. And for large $\eta_i>\eta_c$, two modes can coexist;
whereas for small $\eta$, only electron mode exists.
Fig.\ref{fig:mgk0d_dr_scan_eta}c confirms that for $\eta_i=0.1$,
only one unstable mode exists in the system. The critical gradient
$\eta_c$ is around $2/3$. In this aspect, the ion entropy mode is
similar to ion temperature gradient mode (ITG) in tokamak which is
unstable at larger $\eta_i$; and electron entropy mode is electron
drift mode which can be driven to be unstable by density gradient
alone.

Secondly, we try to understand the relationship of entropy mode and
interchange mode. Usually, the interchange mode and entropy mode are
considered to be two different modes (cf. Ref.\cite{Ricci2006}),
because of their different behaviors in parameters space: For
$k_\parallel=0$ and $k_\perp\to0$, the local interchange mode
$\gamma\to\gamma_0\neq0$; whereas the entropy mode $\gamma\to0$. If
this is the case, there should be a parameter range that both modes
are unstable or the transition from one mode to another in parameter
space should be non-smooth. Entropy mode is considered to be driven
to become unstable by pressure gradient when magnetohydrodynamic
interchange mode is stable. Thus, we scan the gradient parameter
$\epsilon_n$. In Fig.\ref{fig:mgk0d_scan_epsn}a, we find that indeed
an interchange-like mode exists for $\epsilon_n\leq0.2$ and an
entropy-like mode exists for $\epsilon_n\geq0.4$. However, in
Fig.\ref{fig:mgk0d_scan_epsn}c, via scanning of $\epsilon_n$ at
small $k_\perp\rho_i=0.05$, we find the transition between these two
types of mode are smooth, i.e., they are in the same branch. This
also agrees with our previous study
(Fig.\ref{fig:mgk0d_dr_scan_eta}) that for small $\eta_i$, only one
unstable mode exists in the system, either `interchange-like' or
`entropy-like'.

In summary, different numerical approaches can yield the same
solution at convergent limit, but each approach has their own
advantages and disadvantages. DR method is fast but requires good
initial guess and can not give phase space structure directly; IVP
and PIC are slow but can simulate damped mode directly. Matrix
method is the best method for the present study, which can give all
solution in the system and are also fast. However, matrix method can
not treat damped mode correctly, which is probably because damped
mode is the combination of Case-van Kampen modes and not a real
eigenmode (cf. Ref.\cite{Xie2013b} and references in for more
detailed discussions). Combination of multi-approaches can help to
cross check each other and can help to reveal unclear physics in
single approach, as shown by the two examples above.

\section{Extension to 1D}\label{sec:1d}
In this section, we treat the 1D toroidal ion temperature gradient
mode (ITG). The electron is assumed adiabatic $h_e=0$.

We focus on the linearized electrostatic gyrokinetic ITG equation in
ballooning space, ${\bf b}\cdot {\bf
\nabla}=\partial_l=\frac{1}{qR}\frac{\partial}{\partial\theta}$,
where $\theta$ is the ballooning angle coordinate. The gyrokinetic
equation changes to \cite{Dong1992}:
\begin{eqnarray}\label{eq:itg1d0}
    \frac{iv_\parallel}{qR}\frac{\partial}{\partial\theta}h+(\omega-\omega_D)h&=&
      (\omega-\omega_{*T})J_0(\beta)F_M\frac{en_{0i}}{T_i}\phi(\theta),\\
    (1+1/\tau_e)\frac{en_{0i}}{T_i}\phi&=&\int d^3v J_0(\beta)h,
\end{eqnarray}
where $\beta=k_\perp v_\perp/\Omega_{ci}$, $q_i=-q_e=e$,
$b_i=k_\perp^2\rho_{ti}^2$, $\rho_{ti}=v_{ti}/\Omega_{ci}$,
$v_{ti}=(T_i/m_i)^{1/2}$, $\Omega_{ci}=eB/m_ic$, $h$ is
non-adiabatic response of ion,
$\omega_{*T}=\omega_{*i}\{1+\eta_i[v_\perp^2/(2v_{ti}^2)+v_\parallel^2/(2v_{ti}^2)-3/2]\}$,
$\omega_D=2\epsilon_n\omega_{*i}[\cos\theta+s(\theta-\theta_k)\sin\theta](v_\perp^2/2+v_\parallel^2)/(2v_{ti}^2)$,
$\omega_{*i}=-(ck_\theta T_i)/(eBL_n)$, $L_n=-(d\ln n/dr)^{-1}$,
$\eta_i=L_n/L_{T_i}$, $\tau_e=T_e/T_i$, $s=(r/q)(dq/dr)$,
$F_{M}=(\frac{m_i}{2\pi T_i})^{3/2}e^{-m_iv^2/2T_i}$,
$k_\perp^2=k_\theta^2[1+s^2(\theta-\theta_k)^2]$. $\theta_k$ is
ballooning angle parameter. We consider only passing particles, with
also $v_\parallel$ unchanged along the field line.

Normalization: ${\hat \omega}=\omega/\omega_0$, $\omega_0=v_0/r_0$,
$v_0=v_{ti}$, $r_0=R$. ${\hat k}= k\rho_{ti}$, ${\hat v}= v/v_{ti}$.
${\hat\phi}=e\phi/T_i$, ${\hat h}=hv_{ti}^3/n_{0i}$. In the below,
we omit the hat symbol. The normalized equation is
\begin{eqnarray}\label{eq:itg1d}
    \frac{iv_\parallel}{q}\frac{\partial}{\partial\theta}h+(\omega-\omega_D)h&=&
      (\omega-\omega_{*T})J_0(\beta)F_M\phi(\theta),\\\label{eq:itg1d_phi}
    (1+1/\tau_e)\phi&=&\int d^3v J_0(\beta)h,
\end{eqnarray}
where $\beta=k_\perp v_\perp$, $b_i=k_\perp^2$,
$\omega_{*T}=\omega_{*i}[1+\eta_i(v_\perp^2+v_\parallel^2-3)/2]$,
$\omega_D=\omega_{di}[\cos\theta+s(\theta-\theta_k)\sin\theta](v_\perp^2/2+v_\parallel^2)$,
$\omega_{*i}=-k_\theta/\epsilon_n$, $\omega_{di}=-k_\theta$ and
$F_M=(2\pi)^{-3/2}e^{-v^2/2}$, $\int d^3
v=2\pi\int_{-\infty}^{\infty}dv_\parallel\int_{0}^{\infty}v_\perp
dv_\perp$.

\subsection{Integral Dispersion relation}
For completeness, in this subsection, we outline the derivation of
integral equation of the solution to the Eqs.(\ref{eq:itg1d}) and
(\ref{eq:itg1d_phi}). As from Eq.(\ref{eq:itg1d})
\begin{eqnarray*}
  \frac{\partial}{\partial\theta}\Big[h\exp\Big(-i\int^\theta d\theta'(\omega-\omega_D)\frac{q}{v_\parallel}\Big)\Big] =
  -iF_M\\
  \frac{q}{v_\parallel}J_0(\omega-\omega_{*T})
  \delta\phi(\theta)\exp\Big(-i\int^\theta
  d\theta'(\omega-\omega_D)\frac{q}{v_\parallel}\Big),
\end{eqnarray*}
we can obtain
\begin{equation*}
    h^+(\theta)=-i\int_{-\infty}^{\theta}d\theta'F_M\frac{q}{|v_\parallel|}J_0(\omega-\omega_{*T})
  \delta\phi(\theta')\exp(-iI_\theta^{\theta'}),
\end{equation*}
and
\begin{equation*}
    h^-(\theta)=-i\int_{\theta}^{\infty}d\theta'F_M\frac{q}{|v_\parallel|}J_0(\omega-\omega_{*T})
  \delta\phi(\theta')\exp(+iI_\theta^{\theta'}),
\end{equation*}
where
\begin{equation*}
    I_\theta^{\theta'}=\int_{\theta}^{\theta'}d\theta''(\omega-\omega_D)\frac{q}{|v_\parallel|}.
\end{equation*}
Or, we combine the $h^+$ and $h^-$
\begin{eqnarray*}
h(v_\parallel>0)+h(v_\parallel<0)=-i\int_{-\infty}^{\infty}d\theta'
    F_M\frac{q}{|v_\parallel|}\\
    J_0(\beta')(\omega-\omega_{*T})
  \delta\phi(\theta')\exp\Big[i\int_{\theta'}^{\theta}d\theta''(\omega-\omega_D)\frac{qS}{|v_\parallel|}\Big],
\end{eqnarray*}
where $S\equiv sign(\theta-\theta')$. And thus the following
Fredholm integral equation is obtained via substituting the above
solution of $h$ in to the field equation (\ref{eq:itg1d_phi})
\begin{equation}\label{eq:gk_int}
    L\delta\phi(\theta)\equiv(1+\tau)\delta\phi(\theta)-
    \int_0^{\infty}dv_\parallel\int_0^{\infty}dv_\perp\int_{-\infty}^{\infty}d\theta'K_0=0,
\end{equation}
\begin{eqnarray*}
K_0(v_\parallel,v_\perp,\theta,\theta',\omega)\equiv-2\pi iv_\perp\tau
    \frac{q}{|v_\parallel|}(\omega-\omega_{*T})\\
    J(\beta')J(\beta)F_0\delta\phi(\theta')
    \exp\Big[i\int_{\theta'}^{\theta}d\theta''\frac{\omega-\omega_D}{|v_\parallel|}qS\Big],
\end{eqnarray*}
with boundary condition $\delta\phi\to0$ at $\theta\to\pm\infty$,
and
\begin{eqnarray*}
&&\int_{\theta'}^{\theta}d\theta''\frac{\omega-\omega_D}{|v_\parallel|}qS=\frac{\omega}{|v_\parallel|}q|\theta-\theta'|-
\frac{\omega_{di}}{|v_\parallel|}q(v_\perp^2/2+v_\parallel^2)\\
&&~~~~~~S[(1+s)\sin\theta''+s(\theta_k-\theta'')\cos\theta'']\Big|_{\theta'}^{\theta}.
\end{eqnarray*}
The above derivation is similar to Ref.\cite{Romanelli1989} and
\cite{Lu2017}.

In numerical aspect, we use Ritz method via basis functions
expansion
\begin{eqnarray*}
  \delta\phi(\theta) &=& \sum_m\delta\phi_mh_m(\theta-\theta_c), \\
  h_m(\theta) &=&
  \exp[-(c_1\theta)^2/2]H_m(c_1\theta)\sqrt{\frac{c_1}{2^mm!\sqrt{\pi}}},
\end{eqnarray*}
where $c_1$ and $\theta_c$ are adjustable parameters,
\begin{equation*}
    \int_{-\infty}^{\infty}dxh_m(x)h_{m'}(x)=\delta_{mm'}
\end{equation*}
and $H_m$ is the $m$-th Hermite polynomial, i.e.,
\begin{eqnarray*}
  H_0(x)=1,~~H_1(x)=2x, \\
  H_2(x)=4x^2-2,~~H_3(x)=8x^3-12x,~~\cdots
\end{eqnarray*}
Thus, Eq.(\ref{eq:itg1d}) is solved as matrix eigenvalue problem
\begin{equation}\label{eq:mat}
    M_{mm'}\delta\phi_m=0,
\end{equation}
\begin{equation}\label{eq:mlm}
    M_{mm'}=(1+\tau)\delta_{mm'}-\int_{-\infty}^{\infty}d\theta h_{m'}(\theta)
    \int_{0}^{\infty}dv_\parallel\int_{0}^{\infty}dv_\perp\int_{-\infty}^{\infty}d\theta'
    K_0.
\end{equation}
The above approach is also used in FULL code\cite{Rewoldt1982} and
Ref.\cite{Lu2017}, whereas HD7\cite{Dong1992} code use a rectangle
and triangle basis functions.

\subsection{Euler initial value problem}

Define $g(\theta,v_\parallel,v_\perp)\equiv
h(\theta,v_\parallel,v_\perp)-J_0(\beta)F_M(v_\parallel,v_\perp)\phi(\theta)$,
and use $\omega=i\partial_t$, we obtain
\begin{eqnarray}\label{eq:itg1d2}\nonumber
    \partial_tg&=&-\frac{v_\parallel}{q}[\partial_\theta g
    +J_0F_M\partial_\theta\phi+(\partial_\theta J_0)F_M\phi]\\
    &&-i\omega_Dg+i(\omega_{*T}-\omega_D)J_0F_M\phi,\\
    \phi&=&\frac{1}{(1+1/\tau_e-\Gamma_0)}\int d^3v J_0g,
\end{eqnarray}
where $\Gamma_0=I_0(b_i)e^{-b_i}$, and $I_0$ is the modified Bessel
function. And $\partial_\theta J_0=-J_1(\beta)k_\theta v_\perp
s^2(\theta-\theta_k)/\sqrt{1+s^2(\theta-\theta_k)^2}$, since
$J_0'(x)=-J_1(x)$. This approach is similar to GS2
\cite{Kotschenreuther1995} code.

\subsection{Euler eigenvalue problem}
With Euler discretization to eigen matrix and
$\lambda=\partial_t=-i\omega$, we obtain
\begin{eqnarray}\nonumber
    \omega g&=&-i\frac{v_\parallel}{q}[\partial_\theta g
    +J_0F_M\partial_\theta\phi+(\partial_\theta J_0)F_M\phi]\\\label{eq:itg1d3_1}
    &&+\omega_Dg-(\omega_{*T}-\omega_D)J_0F_M\phi,\\\label{eq:itg1d3_2}
    \omega\phi&=&\frac{1}{(1+1/\tau_e-\Gamma_0)}\int d^3v J_0\omega g,
\end{eqnarray}
The $\omega g$ in the integral of Eq.(\ref{eq:itg1d3_2}) is replaced
by the right hand side of Eq.(\ref{eq:itg1d3_1}), and yields
\begin{eqnarray*}
\int d^3v J_0\omega
g=2\pi\int_{-\infty}^{\infty}dv_\parallel\int_{0}^{\infty}v_\perp
dv_\perp
J_0[-i\frac{v_\parallel}{q}\partial_\theta g+\omega_Dg]    \\
    +\{\omega_{di}f_d[(2-b_i)\Gamma_0+b_i\Gamma_1]
    -\omega_{*i}[(1-b_i\eta_i)\Gamma_0+b_i\eta_i\Gamma_1]\}\phi,
\end{eqnarray*}
where $f_d=\cos\theta+s(\theta-\theta_k)\sin\theta$. The eigenvalue
problem is $\omega X=M X$, where
$X=[g(\theta,v_\parallel,v_\perp),\phi(\theta)]$. For the central
difference of $\partial_\theta$, we noticed that high order finite
difference is required, e.g., we find the 4th order coefficients
$(1/12,-2/3,0,2/3,-1/12)$ for $\phi_{j-2,j-1,j,j+1,j+2}$ is much
better than the 2nd order coefficients $(-1/2,0,1/2)$ for
$\phi_{j-1,j,j+1}$.

\subsection{Particle simulation}
The 1D model can also be solved via particle simulation. Defining
the $\delta f$ weight $w=g/F_M$, the gyrokinetic system changes to
\begin{equation}\label{eq:gk_dxi1}
 \frac{d\theta}{dt} = \frac{v_\parallel}{q},
\end{equation}
\begin{eqnarray}\label{eq:gk_dw1}\nonumber
\frac{dw}{dt}&=& -i\omega_{D}w +i(\omega_{*T} - \omega_{D}
  ) J_{0} \phi - \\&&\frac{v_{\parallel}}{q}[J_{0} \partial_{\theta}
  \phi+
  (\partial_{\theta}J_0)
  \phi],
\end{eqnarray}
\begin{equation}\label{eq:gk_ni1}
\Big(1+\frac{1}{\tau_e}-\Gamma_{0}\Big)\phi= \int J_{0} g d^{3} v.
\end{equation}

The simulation steps (here $j=1,2,\cdots, n_p$ is particles index,
and $n_p$ is the total particle number) are:
\begin{itemize}
  \item 1. Loading particles, $\theta$ uniform $rand(np)$, $v_\parallel$
  Maxwellian $randn(np)\times v_{t}$, the perpendicular velocity should be careful $v_\perp=\sqrt{randn(np)^2+randn(np)^2}\times
  v_{t}$;
  \item 2. Calculating $k_\perp v_\perp$, $\omega_{*T}$, $\omega_{D}$ for each particle $j$;
  \item 3. Push particles $\theta_j$,  update $w_j$ for each particle;
  \item 4. Calculating $\phi(\theta)$, with $J_0$ and $\partial_\theta(J_0)$, $\int J_{0}g_i dv^3$;
  \item 5. Back to step 2.
\end{itemize}

This approach is similar to the approach in Ref.\cite{Zhao2002} and
AWECS \cite{Bierwage2008} code.


\begin{figure}
 \centering
  \includegraphics[width=8.5cm]{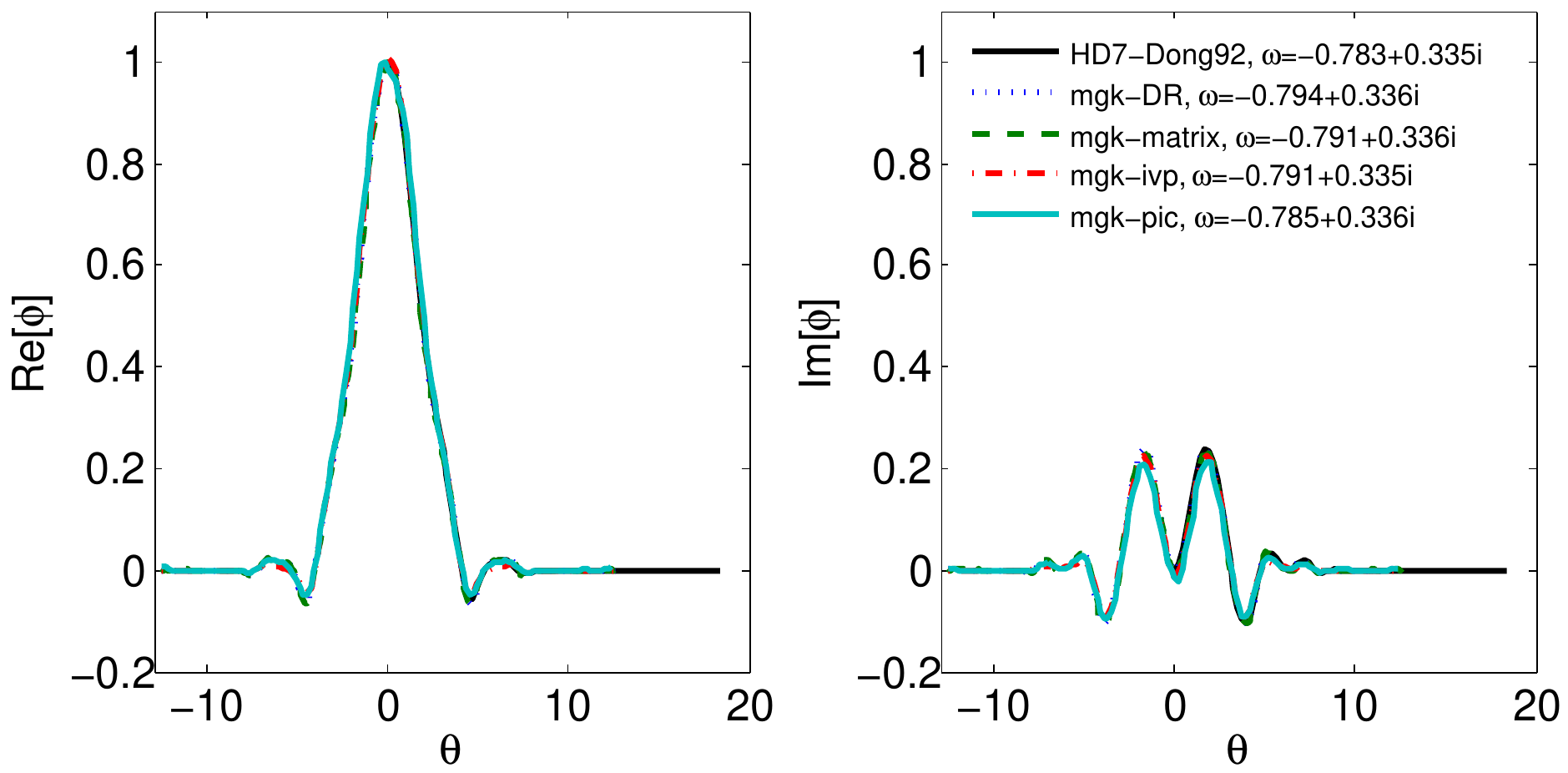}\\
  \caption{Benchmark and comparisons of four approaches with Ref.\cite{Dong1992} ITG solution.
  Both complex frequency and mode structure are close to each other.}\label{fig:mgk1d_dong92}
\end{figure}


\begin{widetext}

\begin{figure}
 \centering
  \includegraphics[width=13.5cm]{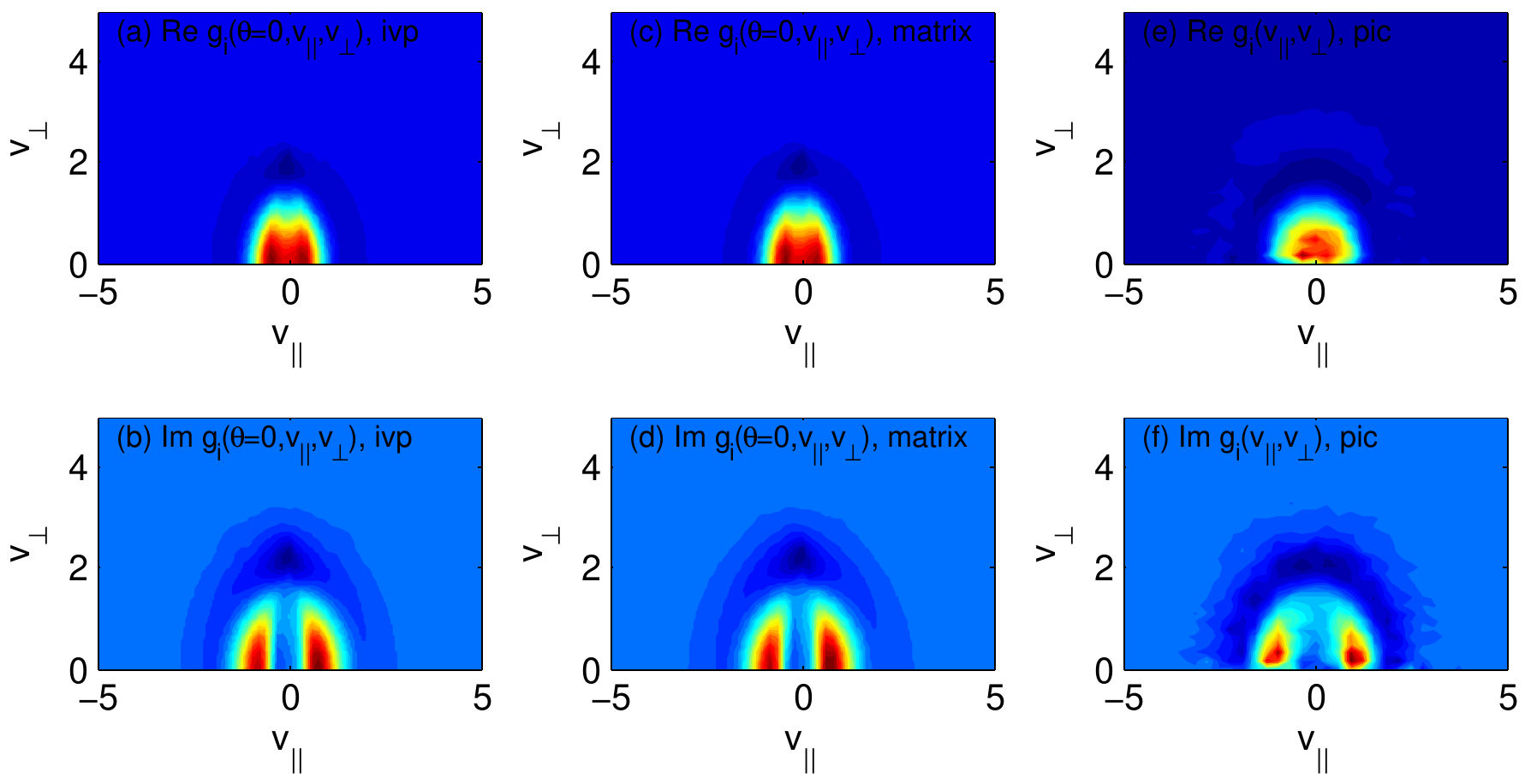}\\
  \caption{Comparisons of $g_i(v_\parallel,v_\perp)$ with the same case of Fig.\ref{fig:mgk1d_dong92}.
  }\label{fig:cmp_gv_dong92}
\end{figure}
\end{widetext}

\begin{figure}
 \centering
  \includegraphics[width=8.5cm]{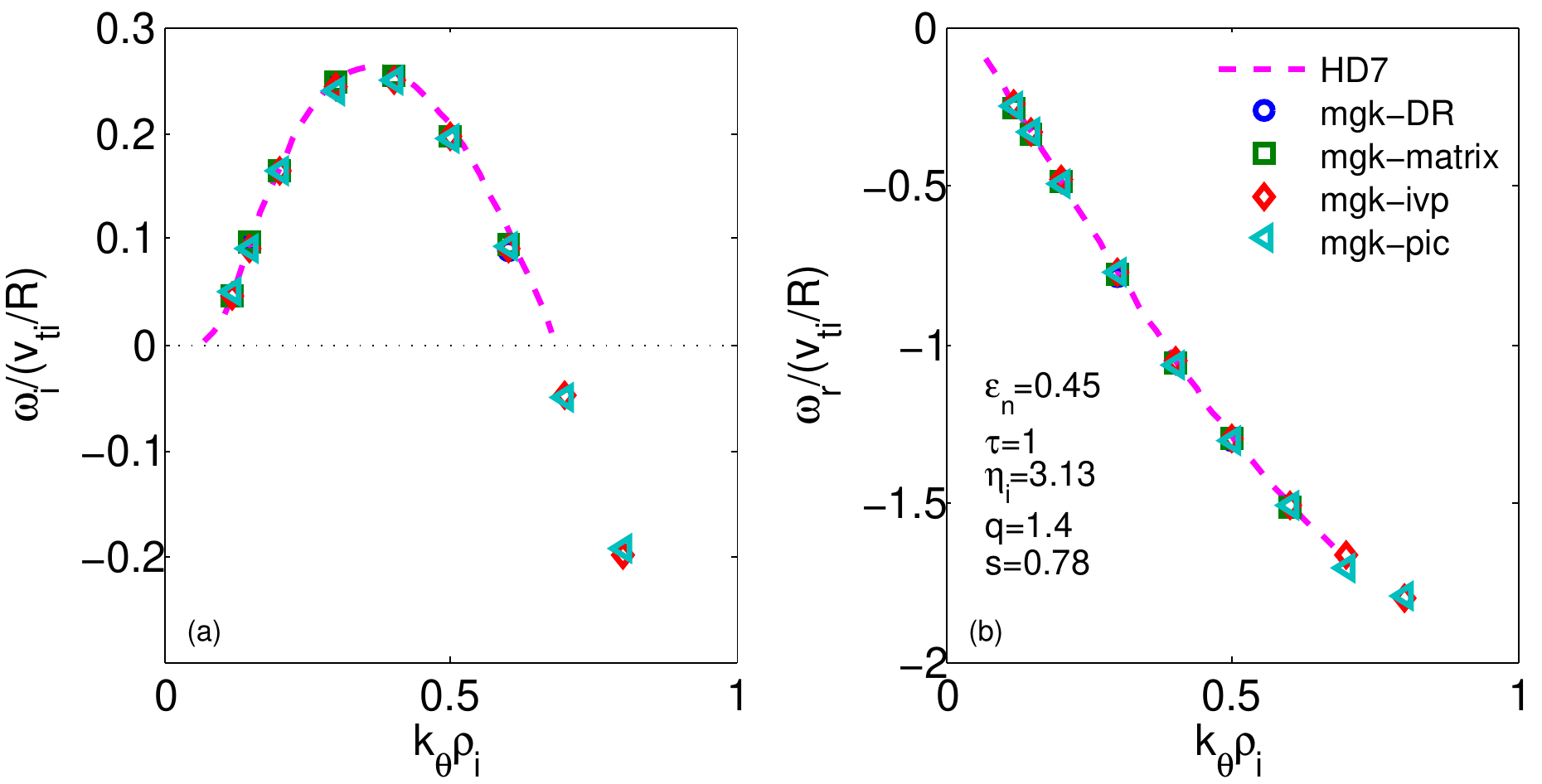}\\
  \caption{Benchmark and comparisons of four approaches with the Cyclone parameters.}\label{fig:mgk1d_cbc}
\end{figure}

\begin{widetext}

\begin{table}[!h]
\tabcolsep 5mm \caption{Typical grids and cpu time for
mgk1d.}\label{tab:mgk1d_1}
\begin{center}
\begin{tabular}{c|cc}
\hline\hline Approach & Grids & Typical cputime\\\hline
DR & $n_\theta=400$, $n_M=25$, $n_{v_\parallel}=31$, $n_{v_\perp}=31$ & $\sim$5min \\
matrix & $n_\theta=200$, $n_{v_\parallel}=32$, $n_{v_\perp}=32$ & $\sim$1min \\
ivp & $n_\theta=128$, $n_{v_\parallel}=64$, $n_{v_\perp}=64$, $\Delta t=0.01$, $n_t=10^3$ & $\sim$30min\\
pic & $n_\theta=128$, $n_p=4\times10^5$, $\Delta t=0.01$, $n_t=10^3$
& $\sim$30min
\\\hline\hline
\end{tabular}
\end{center}
\end{table}
\end{widetext}

\subsection{Benchmark and comparisons}
We benchmark the present solvers with Ref.\cite{Dong1992}, which is
also based on integral method. However, the equation solved in
Ref.\cite{Dong1992} has adopted analytical integral along $v_\perp$
and modified the integral $dv_\parallel$. It is not clear whether
the latter step will affect the final solution. The parameters are:
$k_\perp\rho_i=0.45/\sqrt{2}$, $s=q=\tau=1.0$, $\epsilon_n=0.25$,
$\eta_i=2.5$. The original data in \cite{Dong1992} is
$\omega^{Dong92}=(-0.607+0.258i)\omega_{se}=-0.773+0.328i$. The
newly calculated with fine grids using the same solver HD7 gives
$\omega^{HD7}=(-0.615+0.263i)\omega_{se}=-0.783+0.335i$. The results
are shown in Fig.\ref{fig:mgk1d_dong92}, where we can find that four
approaches and HD7 results agree very well with difference smaller
than $1\%$ for both complex frequency and mode structure.

The velocity space structure $g_i(v_\parallel,v_\perp)$ of three
approaches are shown in Fig.\ref{fig:cmp_gv_dong92}. We can find
that IVP and matrix are almost the same, whereas PIC is also similar
but not exactly the same. This is probably because we gather all the
particles not only just around $\theta=0$ in the PIC method, whereas
in IVP and matrix method we only plot $g_i(\theta=0)$. Another
reason may be the particle noise, since only $n_p=1.6\times10^5$
particles are used. The typical computation grids and cpu time are
shown in \ref{tab:mgk1d_1}. We find in this 1D case, matrix method
is the most fast and accurate approach. And the DR method is not the
fastest method any more which can mainly due to the quadruple
integral. And thus matrix method will be our prior choice.

Next, we benchmark the well known Cyclone case parameters: $s=0.78$,
$q=1.4$, $\tau=1$, $\eta_i=3.13$, $\epsilon_n=0.45$. The results are
shown in Fig.\ref{fig:mgk1d_cbc}. We can find that four approaches
can also agree well. The IVP and PIC method can be more close to the
matrix method when we use more grids or particles. However, we
noticed a slight ($\sim7\%$) discrepancy of the growth rate between
mgk and HD7 at $k_\theta\rho_i\geq0.4$. The reason is not clear yet,
which is probably due to round off error of mgk-matrix, particle
noise of mgk-pic, grid convergence of mgk-dr and mgk-ivp, or the
grid convergence in HD7, especially the velocity space integral of
HD7 using Gauss method which may be difficult to be high accuracy.
We have also tested the MPI parallelized Fortran version of
mgk1d-pic for $k_\theta\rho_i=0.6$, the convergent result with up to
$n_p=6.4\times10^5$ particles is $\omega=-1.51+0.089i$, close to the
one in Fig.\ref{fig:mgk1d_cbc}, whereas $\omega^{\rm
HD7}=-1.50+0.105i$.

\begin{widetext}

\begin{figure}
 \centering
  \includegraphics[width=13.5cm]{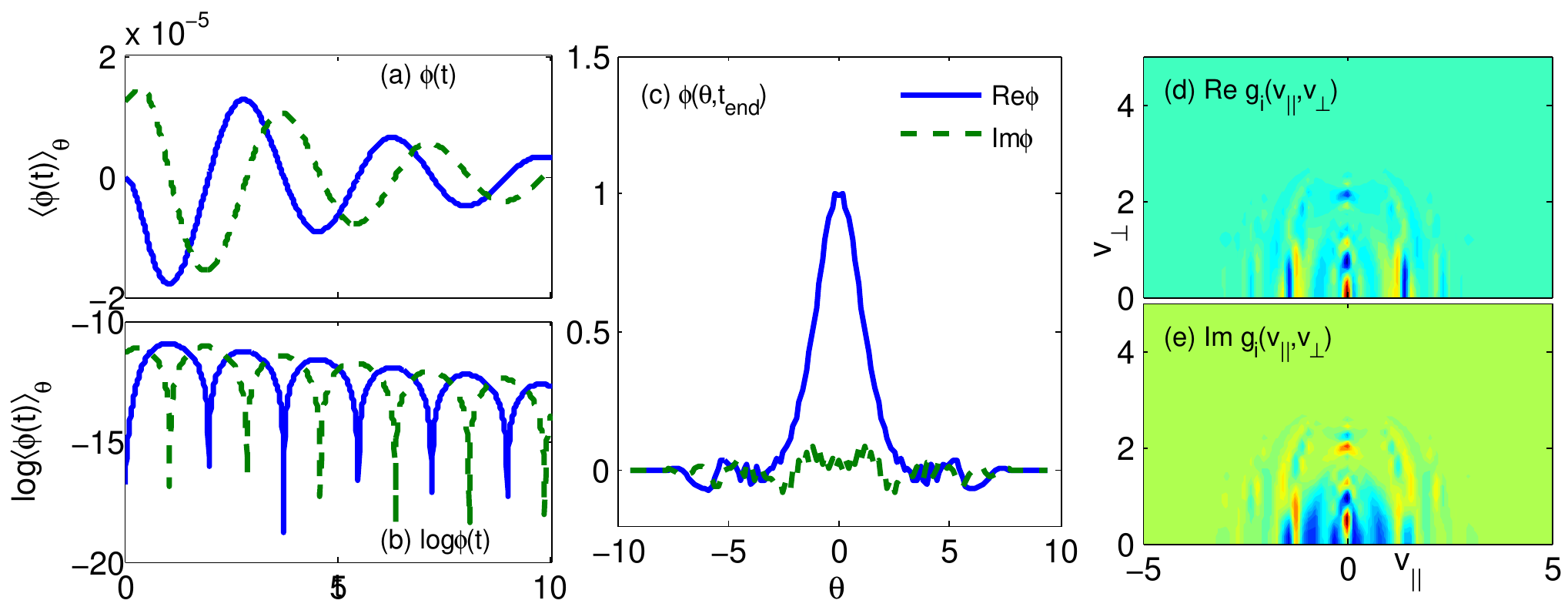}\\
  \caption{For cyclone parameters, the $k_\perp\rho_i=0.8$ mode is
  damped, which can be seen in ivp simulation.
  }\label{fig:mgk1d_ivp_k08_t1000}
\end{figure}
\end{widetext}

The ITG is damped at $k_\theta\rho_i>0.7$ as shown in
Fig.\ref{fig:mgk1d_cbc}a via IVP and PIC method. We further shown a
typical mode structure of this damped mode in
Fig.\ref{fig:mgk1d_ivp_k08_t1000} with $k_\theta\rho_i=0.8$, which
is simulated by IVP. And the PIC method shows similar result. Damped
ITG is studied at Ref.\cite{Sugama1999} via integral method, where
careful treatment of the analytical continuation is required.


\begin{figure}
 \centering
  \includegraphics[width=8.5cm]{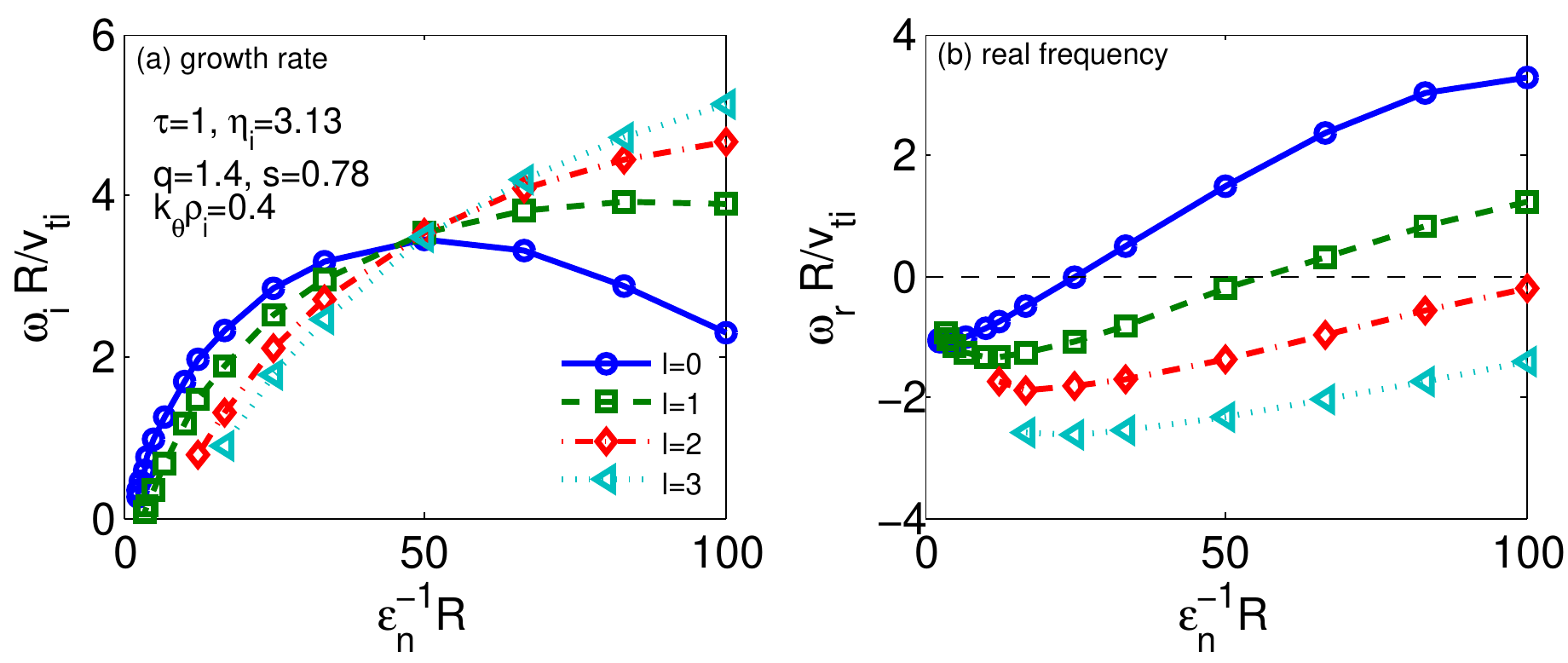}\\
  \caption{At strong gradient, the most unstable ITG mode transits from
  even mode of $l=0$
  ground state to $l\geq1$ high order ITG modes at
  $\epsilon_n^{-1}R\simeq50$. Note also the change of the direction of the real
  frequency from ion direction to electron direction.}\label{fig:cbc_scan_epsn}
\end{figure}

\begin{widetext}

\begin{figure}
 \centering
  \includegraphics[width=14.5cm]{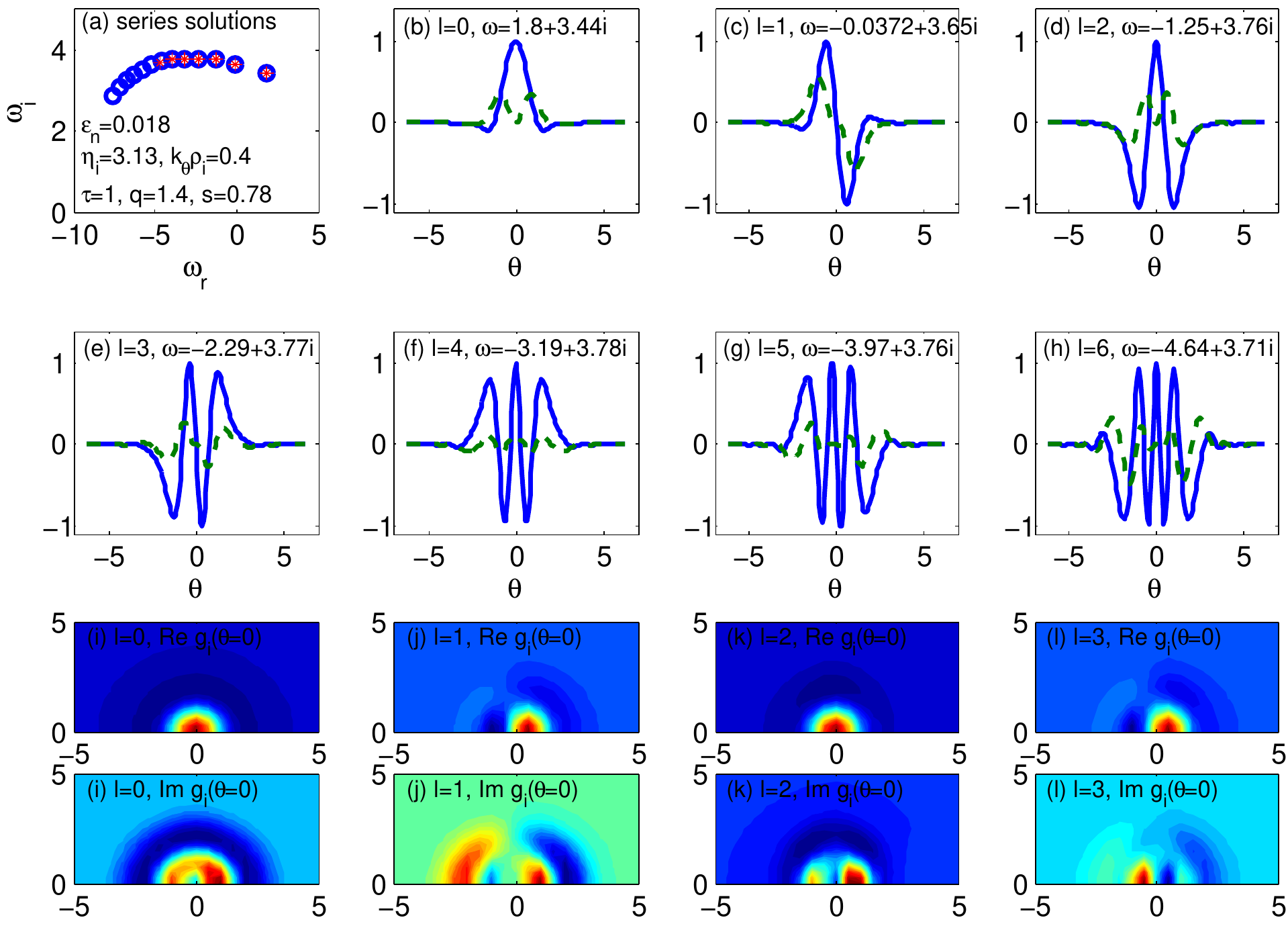}\\
  \caption{For cyclone parameters, $k_\perp\rho_i=0.4$ and $\epsilon_n=0.018$.
  Multi-eigenmodes are shown, where the most unstable modes are around quantum number
  $l\simeq2-5$. Matrix method, which is both fast and accurate under
  this parameter.
  }\label{fig:cbc_multi_eig}
\end{figure}
\end{widetext}

\section{Strong gradient ITGs}\label{sec:gradient}

The above four approaches can have various applications, since we
solve the original gyrokinetic model with no approximation and have
verified that they can yield the same solution. In this work, we
focus on strong gradient, where very interesting new physics are
expected \cite{Xie2017}. Especially, we hope to understand more of
the multi eigenstates at tokamak edge
\cite{Xie2015,Xie2016,Han2017}, where more than one unstable ITGs or
trapped electron modes (TEMs) represent different eigenstates (we
will label them by quantum number `$l$') are found under particular
physical parameters.

We still use the Cyclone parameters but change gradient parameter
$\epsilon_n$, where we find multi-eigenstates coexist, as shown in
Fig.\ref{fig:cbc_scan_epsn}. In this section, we mainly use matrix
method, since IVP and PIC methods are difficult to separate
multi-modes with close growth rate and strongly depend on initial
setups and thus we have not shown their results here. The DR method
requires large grids to converge. And we have checked that the DR
method can converge to the matrix result via increasing grids. We
can find that the $l=0$ ground state ITG mode is no longer the most
unstable one at around $\epsilon_n^{-1}>50$. The growth rate of the
ground state mode increases and then decreases, and the real
frequency of it transits from ion to electron diamagnetic direction
at around $\epsilon_n^{-1}>20$. This is not clearly found by the
systematically parameters scan in Ref.\cite{Han2017} via HD7 code.
One reason is that HD7 requires well initial guess. From the above
result, under Cyclone parameters and $k_\theta\rho_i=0.4$, at the
range $20<\epsilon_n^{-1}<50$, the most unstable ITG mode is on the
ground state but propagates at electron diamagnetic direction. {\it
This new feature tells us that the propagation direction is not a
decisive criterion of the mode for the experimental diagnosis of
turbulence at the edge plasmas.}

In Fig.\ref{fig:cbc_scan_epsn}, we also noticed that ITGs transition
is not a sudden transition, but several modes with similar growth
rates are most unstable at almost the same critical gradient. This
may also explain why the global gyrokinetic PIC simulation in
Ref.\cite{Xie2015} for high order ITG modes does not have clear mode
structures, whereas the TEMs simulation can separate well. This
feature also leads difficulties for IVP and PIC to clearly study the
multi-eigenstates of ITGs.

We show series solutions of ITGs under the same Cyclone parameters
with $\epsilon_n=0.018$ via matrix method in
Fig.\ref{fig:cbc_multi_eig}. Artificial solutions with non-smooth
mode structures and higher order solutions have been removed in
Fig.\ref{fig:cbc_multi_eig}a. We find the most unstable mode under
this parameter is $l\simeq2-5$, and the mode structures of
$\phi(\theta)$ of $l=0-6$ are shown in
Fig.\ref{fig:cbc_multi_eig}b-h. Figure\ref{fig:cbc_multi_eig}i-l
show the velocity space structures of $l=0-3$. We can find that the
mode structures of $g_i(v_\parallel,v_\perp)$ in different
eigenstates $l$ are also different, and especially the odd modes and
even modes are of different types. How to reveal the physical
mechanism and the transitions of the modes through the linear
solution $\omega$, $\phi(\theta)$ and
$g_i(\theta,v_\parallel,v_\perp)$ is out of the scope of present
work.

\section{Summary and Conclusion}\label{sec:summ}
In summary, the present work makes new conclusions regarding ITG
modes in strong gradients along with the development of four new
computational tools for gyro-kinetic simulation of magnetized
plasma. We have developed four independent approaches, based on
integral dispersion relation, Euler initial value simulation, Euler
matrix eigenvalue solution, and Lagrangian particle simulation,
respectively, to study the linear gyrokinetic drift modes. We
identify that these four approaches can yield the same solution and
the differences mainly come from the convergence. These provide a
good tool to study the complete picture of linear drift modes. For
example, using these methods, we find that entropy mode has both
electron and ion branches, and the transition between interchange
mode and entropy mode is a smooth transition and we can consider
them as the same mode. Using matrix method, all solutions in the
system can be revealed and thus we can easily know the distributions
and transitions of the modes in the system. The multi-eigenstates of
gyrokinetic ITG can be seen such clearly for the first time, via
matrix method. Most importantly, we find that the propagation
direction is not a decisive criterion for the experimental diagnosis
of turbulent mode at the edge plasmas, i.e., the most unstable ITG
can propagate at electron direction!

In conclusion, this work, providing four tools, can have many
potential applications in the future to reveal different aspects of
linear kinetic mirco-instabilities. Examples of applications to
strong gradient to reveal new physics are shown for both Z-pinch
entropy mode and tokamak ITGs. Electromagnetic, collision and
trapped particle effects can be added in principle.  The performance
comparisons in the present work is rather crude, which is mainly due
to that it is difficult to determined the computation precision and
thus it may not be fair for the comparisons. Both IVP and PIC
methods scales linearly to the grids/particle numbers. The
performance of the iterative solver of matrix and DR methods depends
on both grid numbers and initial guess. And also, optimizations of
each codes to speed up are possible and can be considered as future
works. We should also emphasize that although the above four
approaches have outlined the major linear methods, the
implementations can be varying, examples including
GENE\cite{Roman2010}, GYRO\cite{Belli2010}, LIGKA\cite{Lauber2007},
etc.

\acknowledgments The authors would like to thank M. K. Han for
providing the benchmark data of HD7 code. Discussions with J. Q.
Dong are also acknowledged. The work was supported by the China
Postdoctoral Science Foundation No. 2016M590008, Natural Science
Foundation of China under Grant No. 11675007 and 11605186, and the
ITER-China Grant No. 2013GB112006.

\appendix
\section{Useful integrals}
Some useful integrals
\begin{eqnarray*}
  &&\int_0^{\infty}x^{2n}e^{-x^2/a^2}dx = \frac{a^{2n+1}(2n-1)!!}{2^{n+1}}\sqrt{\pi}, \\
  &&\int_0^{\infty}x^{2n+1}e^{-x^2/a^2}dx = \frac{n!}{2}a^{2n+2}, \\
  &&\int_0^{\infty}xe^{-x^2/2}J_p^2(\sqrt{b_s}x)dx = e^{-b_s}I_p(b_s),\\
  &&\int_0^{\infty}x^3 e^{-x^2/2}J_p^2(\sqrt{b_s}x)dx =
  2e^{-b_s}[\\&&~~~~~~~~~~(1-b_s+p)I_p(b_s)+b_sI_{p+1}(b_s)],
\end{eqnarray*}
where $J_p$ and $I_p$ are Bessel and modified Bessel functions,
respectively.

\end{document}